\documentclass[a4paper,
               biblatex,     
               ]{jacow}
\usepackage{pdfpages,multirow,ragged2e} %
\makeatletter%
	\ifboolexpr{bool{xetex}}
	 {\renewcommand{\Gin@extensions}{.pdf,%
	                    .png,.jpg,.bmp,.pict,.tif,.psd,.mac,.sga,.tga,.gif,%
	                    .eps,.ps,%
	                    }}{}
\makeatother

\ifboolexpr{bool{xetex} or bool{luatex}} 
 {}                                      
 {\usepackage[utf8]{inputenc}}           

\usepackage[USenglish]{babel}

\usepackage{subcaption}

\ifboolexpr{bool{jacowbiblatex}}%
 {%
  \addbibresource{TUPS013.bib}
 }{}
\begin{document}

\title{Ion-motion simulations of a  \\ plasma-wakefield experiment at FLASHForward}

\author{D. Kalvik\thanks{daniel.kalvik@fys.uio.no}, P. Drobniak, F. Pe\~{n}a\textsuperscript{1}, C. A. Lindstr{\o}m, \\ 
    Department of Physics, University of Oslo, Oslo, Norway \\
        J. Beinortaite, L. Boulton, P. Caminal
        J. Garland, G. Loisch,\\
        J. B. Svensson, M. Thévenet, S. Wesch, J. Wood, \\
        Deutsches Elektronen-Synchrotron, Hamburg, Germany \\
        J. Osterhoff, Lawrence Berkeley National Laboratory, Berkeley, USA \\
        R. D'Arcy, John Adams Institute, Department of Physics, University of Oxford, Oxford, UK\\
        S. Diederichs, CERN, Geneva, Switzerland\\
		\textsuperscript{1}also at Ludwig-Maximilians-Universität München, Munich, Germany}
	
\maketitle

\begin{abstract}
   In plasma-based acceleration, an ultra-relativistic particle bunch---or an intense laser beam---is used to expel electrons from its propagation path, forming a wake that is devoid of electrons. The ions, being significantly more massive, are often assumed to be stationary. However, both theory and simulations suggest that any sufficiently dense electron bunch can trigger ion motion, and its effect must be taken into account. We simulate beam-driven plasma wakefields to identify key features---such as longitudinally dependent emittance growth---that could be observed in an experiment using plasma and beam parameters from the FLASHForward facility at DESY.
\end{abstract}

\section{Introduction}
Plasma-acceleration is a method of accelerating electrons on a significantly shorter length scale than conventional RF-accelerators \cite{TajimaDawson, chen, Ruth:157249}. By injecting a high-intensity laser or an ultra-relativistic charged particle beam---referred to as a \textit{driver}---into a plasma, the longitudinal electric field within the plasma can reach the order of tens of gigavolts per meter \cite{LeemansW.P2006Gebf, LeemansWP2014Mebf, ClaytonChristopherE2007Edo4}. In the most extreme case, the driver completely expels the plasma electrons from its path---creating a bubble void of plasma electrons. The ions however, are often considered stationary due to their mass.

As shown in Refs.~\cite{Rosenzweig, Lee, Gholizadeh}, intense electron bunches are capable of inducing non-negligible ion motion. This ion motion will induce non-linear focusing forces that can cause the emittance of the witness to increase \cite{An}. The preservation of emittance is critical in applications within high-energy physics, such as for free electron lasers and linear colliders. It is therefore of interest to measure and quantify the effects of ion motion experimentally.

In this paper, we show a possible experiment to measure the effects of ion motion; we show the expected results---obtained through simulations---and what would be the diagnostic setup. The simulations are based on the beam and plasma parameters available at the FLASHForward facility at DESY \cite{Darcy}.

\section{Theory}
The amount of ion motion for a round beam can be approximated by the equation given in Ref.~\cite{Rosenzweig},
\begin{equation}
    \Delta \phi \simeq \sqrt{\frac{2\pi Z r_a \sigma_z N_b}{A\varepsilon_{n,x}}}\left(r_en_0\gamma\right)^{1/4}.
    \label{eq:ion_motion}
\end{equation}
The parameters of the equation are the phase-advance, $\Delta \phi$; the net charge of the plasma ion species, $Z$; the classical radius of a singly charged ion of mass $1$ amu, $r_a$; the rms bunch-length, $\sigma_z$; the number of bunch-electrons, $N_b$; the atomic mass of the ion-species (in amu), $A$; the normalized horizontal emittance, $\varepsilon_{n,x}$; the classical electron radius, $r_e$; the unperturbed ion/electron density, $n_0$, and the beam Lorentz-factor, $\gamma$. The phase-advance is a measure of how much the ions oscillate while under the focusing field of the bunch---assumed to be cylindrically symmetric. Maximum ion motion occurs after a phase-advance of $\frac{\pi}{2}$, leading to a full collapse of the ion column.

The effects of the ion motion grow towards the back of the bunch---corresponding to lower values in the co-moving frame of the bubble, $z$---as it is where the ions have spent the most time within its fields. If we insert a single driver into the plasma, and make it long enough to fill most of the bubble, we can correlate the higher energies with the back of the bunch. Measuring the emittance as a function of energy towards the back of the bunch should therefore give us an indicator of ion motion.

\section{Simulations}
In this section, we present the results obtained in the simulation study. To simulate the plasma-acceleration, we used HiPACE++ \cite{Hipace}. 
For the simulations, we use a constant plasma density of \SI{2e15}{cm^{-3}} (up/down-ramps \cite{uniform-ramp}), and \(\SI{1.2e16}{cm^{-3}}\) (flattop). The flattop length is \(\SI{40}{mm}\), and the ramp lengths are \SI{12}{mm} each. The beam has an energy of \SI{1}{GeV}, emittances of (\SI{1.5}{mm\,mrad} (horizontal) and \(\SI{2.5}{mm\,mrad}\) (vertical), a charge of \SI{-0.75}{nC}, and a relative energy spread of \SI{0.5}{\%}.
The $\beta$-functions are down-ramped to five times the matched value in the flattop region, from an initial value that is 30 times larger than the matched value. The current profile is a realistic non-Gaussian profile based on experiments \cite{Darcy, Lindstrom} with an rms bunch-length of \(\SI{71}{\mu m}\) and a peak current of \(\SI{1.1}{kA}\).
The simulation box uses mesh refinement and has two regions. In the flattop, the larger region covers a length in $z$ of \SI{815.0}{$\mu$m} (\SI{877} cells) and a width in both $x$ and $y$ of \SI{950.6}{$\mu$m} (\SI{1023}{}$\times$\SI{1023} cells), 
and a central region with the same length and resolution in $z$ but with \SI{16}{} times the resolution in $x$ and $y$ ($\pm$ \SI{6.1}{$\mu$m}, \SI{511}{}$\times$\SI{511} cells).
In the ramps, the transverse size and resolution are all multiplied by a factor \SI{2.45}{}.
The number of macro-particles (constant weight) is \SI{8e6}{}.

For the spectrometer simulations, we used ImpactX \cite{huebl}. The beam was obtained from the output of the HiPACE++ simulation. The spectrometer consists of 5 quadrupoles, a dipole, and a spectrometer screen. The first two quadrupoles have the same field gradient, the same goes for the next two, while the last has a third, different gradient. The gradient for the first two quadrupoles is \SI{29.55}{T\per m}, for the next two it is \SI{-40.62}{T\per m}, and for the final it is \SI{43.92}{T\per m}. The lengths of all the quadrupoles are \SI{0.1137}{m}. The dipole has a magnetic field of \SI{-0.28}{T} and a length of \SI{1.07}{m}.
The distances between the elements (starting at the plasma) are
\SI{0.66}{m}, \SI{0.27}{m}, \SI{0.26}{m}, \SI{0.27}{m}, \SI{0.38}{m}, \SI{3.9}{m} and \SI{1.38}{m}.

In running the simulations, we have used the Advanced Beginning-to-End Linac (ABEL) simulation framework \cite{ben}. ABEL combines different codes, which allows for agile design and simulations of various beamline elements.
\subsection{Ion Motion}
\begin{figure}[!tbh]
    \centering
    \includegraphics[width=0.9\linewidth]{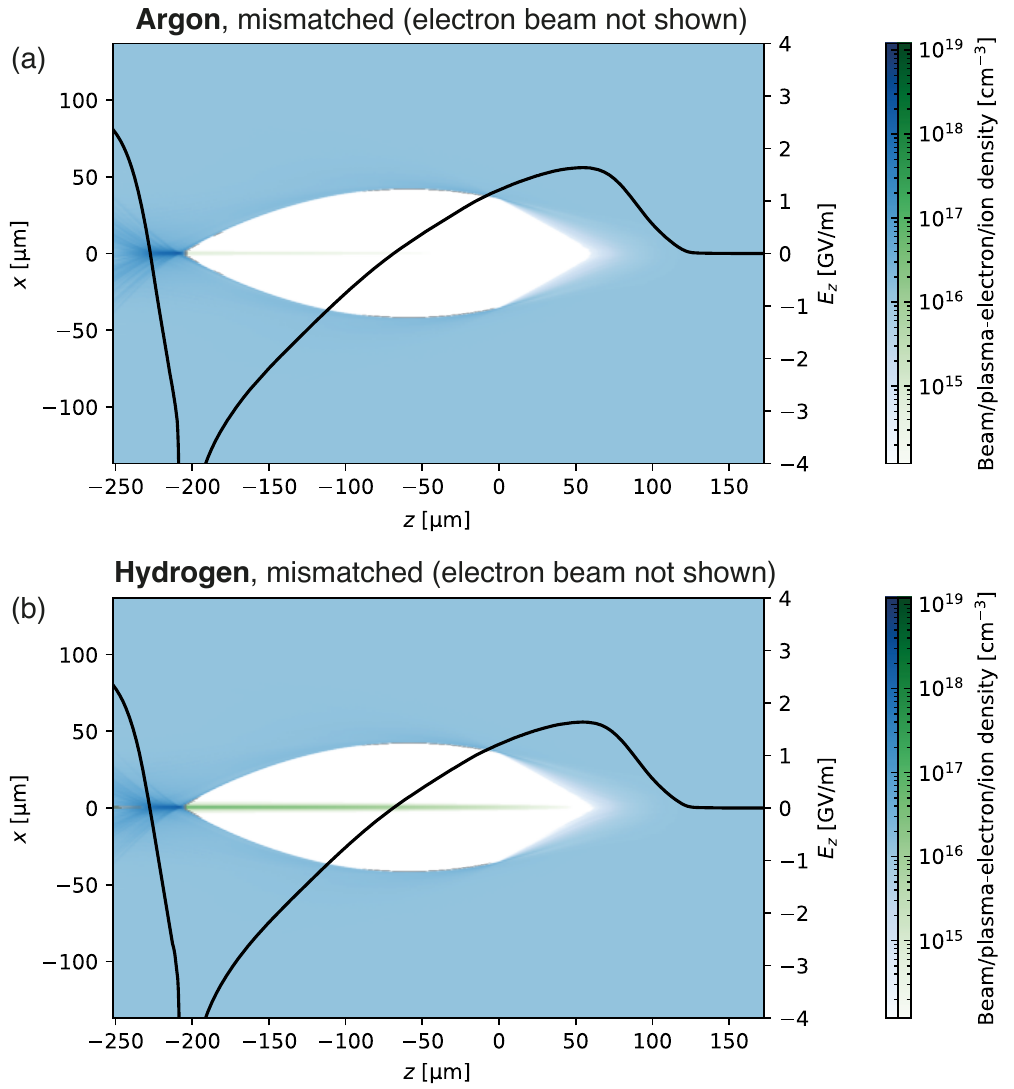}
    \caption{Plasma density of plasma electrons (blue colormap) and plasma ions (green colormap) in (a) argon and (b) hydrogen, without showing the beam density. A strong on-axis ion spike is visible in hydrogen.}
    \label{fig:density}
\end{figure}

For the stated parameters, equation \ref{eq:ion_motion} predicts a phase-advance through the beam, $\Delta\phi=\SI{0.66}{}$ for hydrogen, and $\Delta\phi=\SI{0.10}{}$ for argon. Since the equation is derived for round beams we use $\varepsilon_{n,x}=\sqrt{\varepsilon_{n,x}\varepsilon_{n,y}}$. These parameters do not correspond to a complete collapse of the ion column, even for hydrogen, but we still see a significant difference in the predicted amount of ion motion. Figure \ref{fig:density} shows the plasma density for both electrons and ions in argon and hydrogen. The presence of ion motion is clearly more prevalent in hydrogen compared to argon.

\subsection{Longitudinal phase space}
\begin{figure}[!tbh]
    \centering
    \includegraphics[width=1\linewidth]{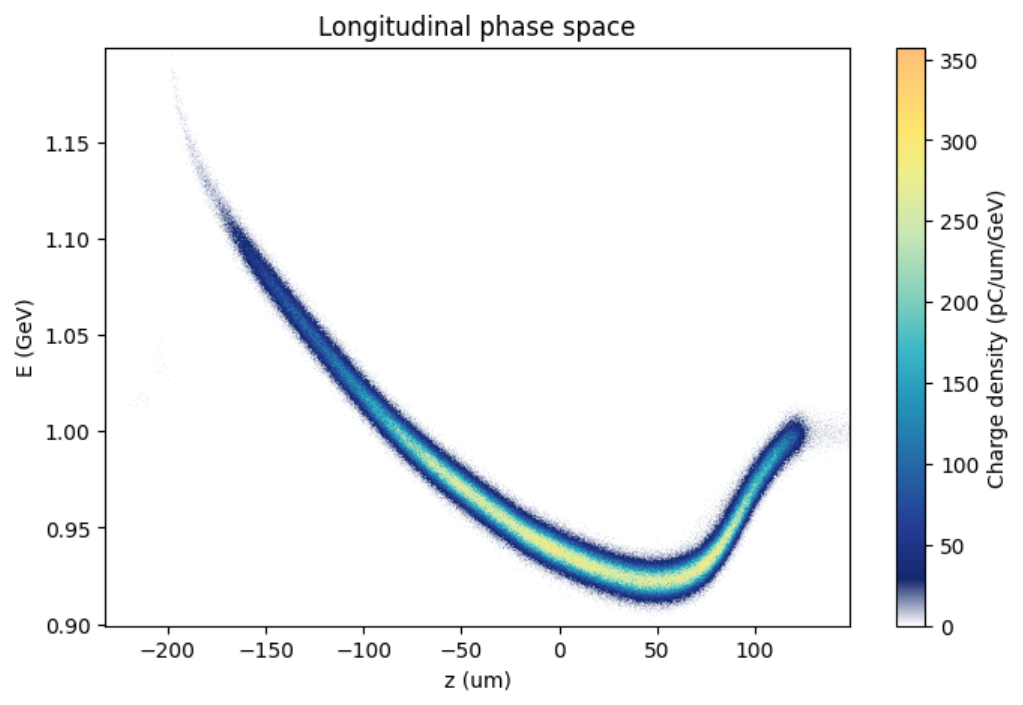}
    \caption{The simulated longitudinal phase space of an electron beam after traversing a plasma. This resulting phase space will generally be similar for both hydrogen- and argon-based acceleration.}
    \label{fig:lps}
\end{figure}

In order to correlate energy with the longitudinal position along the beam, we must verify that the energy of the particles is indeed increasing with longitudinal position within the beam (for the accelerated part of the beam). The longitudinal phase space of the beams after traversing the plasma is shown in Fig.~\ref{fig:lps}. Above \SI{1.02}{GeV} the only particles are the ones in the back of the bubble. Beyond this point, we can be confident in correlating energy with the longitudinal position of the beam.
\subsection{Horizontal emittance}
\label{section:horizontal_emittance}
\begin{figure}[!tbh]
    \centering
    \includegraphics[width=0.9\linewidth]{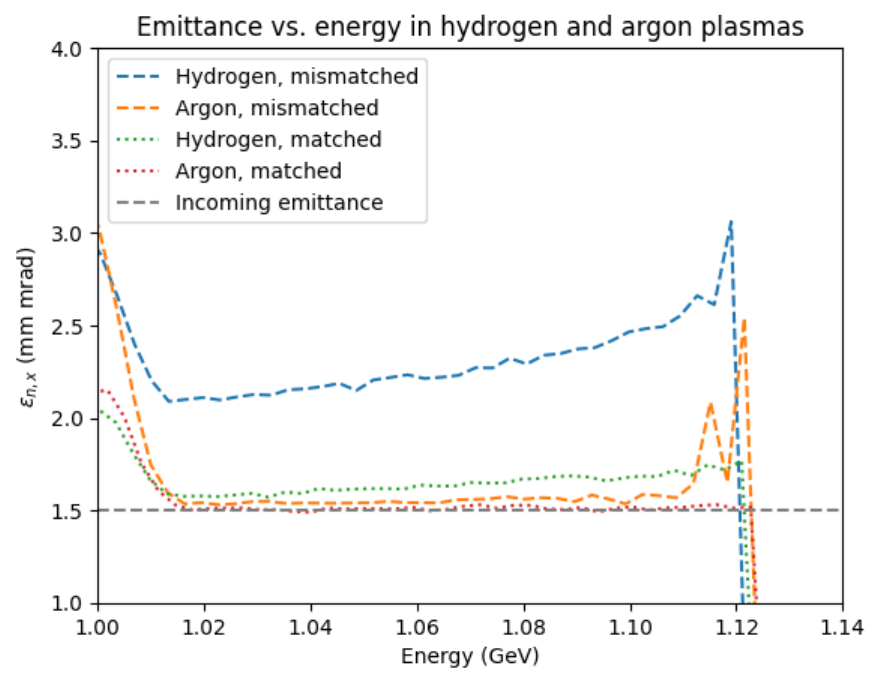}
    \caption{Emittance evolution along both matched and mismatched beams in hydrogen and argon plasmas. Energies above \SI{1.02}{GeV} correlate with lower values of $z$ (see Fig.~\ref{fig:lps}).}
    \label{fig:emittances}
\end{figure}
Figure \ref{fig:emittances} shows the emittance evolution---post plasma-acceleration---for both a mismatched and matched beam, in both argon and hydrogen. In both the matched and mismatched cases, the emittance evolution in hydrogen stands out from that in argon. In the case of a matched beam, the difference is subtle ($\sim$\SI{10}{\%}), and therefore likely to be difficult to measure. The difference is much less subtle in the case of a mismatched beam, as the emittance is seen growing substantially more in hydrogen, as opposed to argon. The reason for this is that the particles will have a larger oscillation amplitude, which will in turn make them see more of the nonlinear focusing induced by the ion motion \cite{Benedetti, An}.

\subsection{Spectrometer imaging}
\begin{figure}
    \centering
    \includegraphics[width=0.9\linewidth]{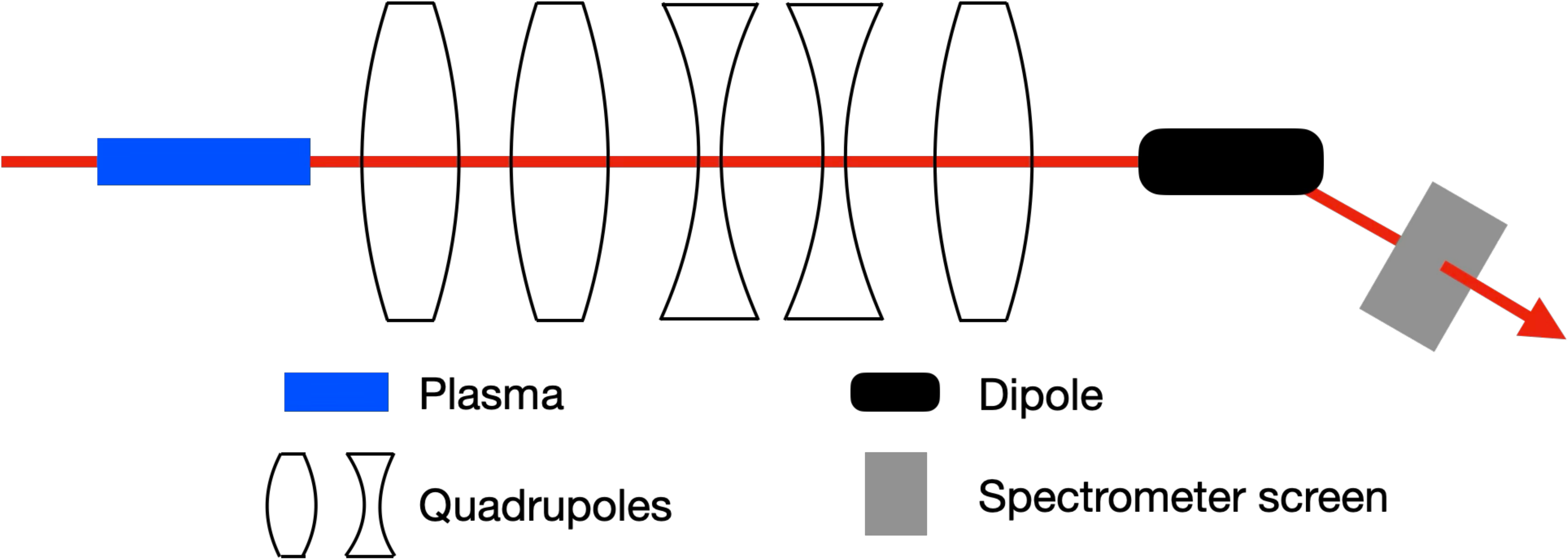}
    \caption{The imaging spectrometer setup. The quadrupoles captures and focuses the beam, the dipole bends the beam in the vertical plane, and the spectrometer screen captures the transverse profile of the beam.}
    \label{fig:FF diagnostics}
\end{figure}

\begin{figure}
    \centering
    \includegraphics[width=0.8\linewidth]{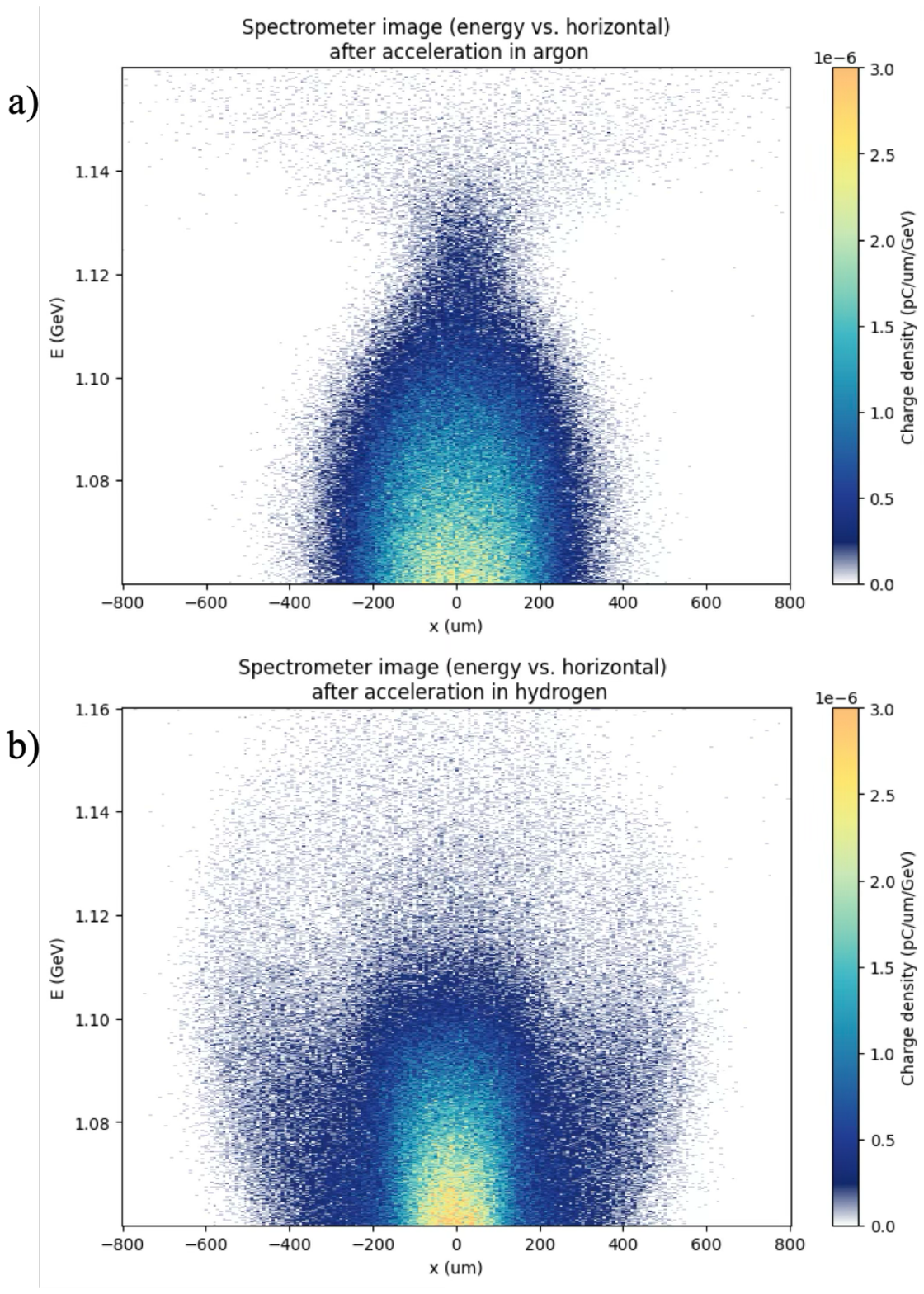}
    \caption{Simulated spectrometer images for (a): Argon and (b): Hydrogen. There is a clear difference in both size and shape between a beam accelerated in hydrogen-plasma, and one accelerated in argon-plasma. The beam is mismatched in both plots.}
    \label{fig:Spec_both}
\end{figure}

To verify Fig.~\ref{fig:emittances} experimentally, one can perform an object-plane scan using an imaging spectrometer \cite{Lindstrom}. We simulate a spectrometer setup similar to that of FLASHForward at DESY---sketched in Fig.~\ref{fig:FF diagnostics}. The beam is focused by the quadrupoles and bent vertically towards the spectrometer screen by the dipole. The screen captures the transverse profile of the dispersed beam, which allows for measuring the energy spectrum in the vertical plane, and the transverse phase space in the horizontal plane.
Figure \ref{fig:Spec_both} shows a single spectrometer image, point-to-point imaged at \SI{1}{GeV}. The same exact parameters have been used in both images, except for the choice of ion species. A distinct feature of these images is the shape of the beam. In the argon plasma, the beam has maintained its Gaussian shape in the horizontal plane, while in the hydrogen plasma, this is not the case.

\section{Conclusion}
We have simulated a beam in both hydrogen and argon plasmas, and simulated the resulting beam through a spectrometer. We have covered two distinct features of ion motion, namely emittance growth, and a change in transverse particle distribution. The emittance growth in hydrogen is significantly larger than in argon, and is therefore a clear sign of ion motion. In addition, the simulated spectrometer images show a non-Gaussian shape in the case of hydrogen-based acceleration, whereas the argon-simulated beam maintained its Gaussian shape. The results are promising for the prospects of observing ion motion induced emittance growth at FLASHForward.

\section{ACKNOWLEDGEMENTS}
This work is funded by the European Research Council (ERC Grant Agreement No. 101116161). We acknowledge Sigma2 - the National Infrastructure for High-Performance Computing and Data Storage in Norway for awarding this project access to the LUMI supercomputer, owned by the EuroHPC Joint Undertaking, hosted by CSC (Finland) and the LUMI consortium. This work was supported by the Director, Office of Science, Office of High Energy Physics, of the U.S. Department of Energy, under Contract No.~DE-AC0205CH11231.
%
%
\ifboolexpr{bool{jacowbiblatex}}%
	{\printbibliography}%

@inproceedings{huebl,
    author = {A. Huebl and others},
    title = {{Next Generation Computational Tools for the Modeling and Design of Particle Accelerators at Exascale}},
    booktitle = {Proc. NAPAC'22},
    pages = {302--306},
    paper = {TUYE2},
    venue = {Albuquerque, NM, USA},
    series = {North American Particle Accelerator Conference},
    publisher = {JACoW Publishing, Geneva, Switzerland},
    month = {11},
    year = {2022},
    issn = {2673-7000},
    isbn = {978-3-95-450232-5},
    doi = {10.18429/JACoW-NAPAC2022-TUYE2},
    url = {https://jacow.org/napac2022/papers/TUYE2.pdf},
    language = {english}
}

@article{Hipace,
    author = {Diederichs, S. and Benedetti, C. and Huebl, A. and Lehe, R. and Myers, A. and Sinn, A. and Vay, J.-L. and Zhang, W. and Thévenet, M.},
    title = {HiPACE++: A portable, 3D quasi-static particle-in-cell code},
    journal = {Computer Physics Communications},
    volume = {278},
    pages = {108421},
    year = {2022},
    doi = {10.1016/j.cpc.2022.108421},
    url = {https://doi.org/10.1016/j.cpc.2022.108421}
}

@article{Rosenzweig,
  title = {Effects of Ion Motion in Intense Beam-Driven Plasma Wakefield Accelerators},
  author = {Rosenzweig, J. B. and Cook, A. M. and Scott, A. and Thompson, M. C. and Yoder, R. B.},
  journal = {Phys. Rev. Lett.},
  volume = {95},
  issue = {19},
  pages = {195002},
  numpages = {4},
  year = {2005},
  month = {Oct},
  publisher = {American Physical Society},
  doi = {10.1103/PhysRevLett.95.195002},
  url = {https://link.aps.org/doi/10.1103/PhysRevLett.95.195002}
}

@article{LeemansW.P2006Gebf,
address = {LONDON},
author = {Leemans, W. P and Esarey, E and Nagler, B and Gonsalves, A. J and Tóth, Cs and Nakamura, K and Geddes, C. G. R and Schroeder, C. B and Hooker, S. M},
issn = {1745-2473},
journal = {Nature physics},
pages = {696--699},
volume = {2},
publisher = {Springer Nature},
number = {10},
year = {2006},
title = {GeV electron beams from a centimetre-scale accelerator},
}

@article{ClaytonChristopherE2007Edo4,
address = {LONDON},
author = {Clayton, Christopher E and Decker, Franz-Josef and Walz, Dieter and Blumenfeld, Ian and Ischebeck, Rasmus and Siemann, Robert H and Oz, Erdem and Zhou, Miaomiao and Katsouleas, Thomas and Kirby, Neil and Marsh, Kenneth A and Huang, Chengkun and Joshi, Chandrashekhar and Muggli, Patric and Lu, Wei and Iverson, Richard and Mori, Warren B and Hogan, Mark J},
issn = {0028-0836},
journal = {Nature},
pages = {741--744},
volume = {445},
publisher = {Springer Nature},
number = {7129},
year = {2007},
title = {Energy doubling of 42 GeV electrons in a metre-scale plasma wakefield accelerator},
}

@article{LeemansWP2014Mebf,
address = {COLLEGE PK},
author = {Leemans, W. P. and Gonsalves, A. J. and Mao, H. S. and Nakamura, K. and Benedetti, C. and Schroeder, C. B. and Tóth, Cs and Daniels, J. and Mittelberger, D. E. and Bulanov, S. S. and Vay, J. L. and Geddes, C. G.R. and Esarey, E.},
issn = {0031-9007},
journal = {Physical review letters},
pages = {245002--245002},
volume = {113},
publisher = {Amer Physical Soc},
number = {24},
year = {2014},
title = {Multi-Gev electron beams from capillary-discharge-guided subpetawatt laser pulses in the self-trapping regime},
}

@article{TajimaDawson,
  title = {Laser Electron Accelerator},
  author = {Tajima, T. and Dawson, J. M.},
  journal = {Phys. Rev. Lett.},
  volume = {43},
  issue = {4},
  pages = {267--270},
  numpages = {0},
  year = {1979},
  month = {Jul},
  publisher = {American Physical Society},
  doi = {10.1103/PhysRevLett.43.267},
  url = {https://link.aps.org/doi/10.1103/PhysRevLett.43.267}
}

@article{chen,
  title = {Acceleration of Electrons by the Interaction of a Bunched Electron Beam with a Plasma},
  author = {Chen, Pisin and Dawson, J. M. and Huff, Robert W. and Katsouleas, T.},
  journal = {Phys. Rev. Lett.},
  volume = {54},
  issue = {7},
  pages = {693--696},
  numpages = {0},
  year = {1985},
  month = {Feb},
  publisher = {American Physical Society},
  doi = {10.1103/PhysRevLett.54.693},
  url = {https://link.aps.org/doi/10.1103/PhysRevLett.54.693}
}

@article{Ruth:157249,
      author        = "Ruth, Ronald D and Chao, A W and Morton, P L and Wilson, P
                       B",
      title         = "{A plasma wake field accelerator}",
      reportNumber  = "SLAC-PUB-3374",
      journal       = "Part. Accel.",
      volume        = "17",
      pages         = "171",
      year          = "1984",
}

@article{Benedetti,
    author = {Benedetti, C. and Mehrling, T. J. and Schroeder, C. B. and Geddes, C. G. R. and Esarey, E.},
    title = {Adiabatic matching of particle bunches in a plasma-based accelerator in the presence of ion motion},
    journal = {Physics of Plasmas},
    volume = {28},
    number = {5},
    pages = {053102},
    year = {2021},
    month = {05},
    issn = {1070-664X},
    doi = {10.1063/5.0043847},
    url = {https://doi.org/10.1063/5.0043847},
    eprint = {https://pubs.aip.org/aip/pop/article-pdf/doi/10.1063/5.0043847/13920677/053102\_1\_online.pdf},
}

@article{An,
  title = {Ion Motion Induced Emittance Growth of Matched Electron Beams in Plasma Wakefields},
  author = {An, Weiming and Lu, Wei and Huang, Chengkun and Xu, Xinlu and Hogan, Mark J. and Joshi, Chan and Mori, Warren B.},
  journal = {Phys. Rev. Lett.},
  volume = {118},
  issue = {24},
  pages = {244801},
  numpages = {5},
  year = {2017},
  month = {Jun},
  publisher = {American Physical Society},
  doi = {10.1103/PhysRevLett.118.244801},
  url = {https://link.aps.org/doi/10.1103/PhysRevLett.118.244801}
}

@article{Darcy,
issn = {1364-503X},
journal = {Philosophical transactions of the Royal Society of London. Series A: Mathematical, physical, and engineering sciences},
pages = {20180392--20180392},
volume = {377},
publisher = {The Royal Society Publishing},
number = {2151},
year = {2019},
title = {FLASHForward: plasma wakefield accelerator science for high-average-power applications},
copyright = {2019 The Author(s) 2019},
language = {eng},
address = {England},
author = {D'Arcy, R. and Aschikhin, A. and Bohlen, S. and Boyle, G. and Brümmer, T. and Chappell, J. and Diederichs, S. and Foster, B. and Garland, M. J. and Goldberg, L. and Gonzalez, P. and Karstensen, S. and Knetsch, A. and Kuang, P. and Libov, V. and Ludwig, K. and Martinez de la Ossa, A. and Marutzky, F. and Meisel, M. and Mehrling, T. J. and Niknejadi, P. and Põder, K. and Pourmoussavi, P. and Quast, M. and Röckemann, J. -H. and Schaper, L. and Schmidt, B. and Schröder, S. and Schwinkendorf, J. -P. and Sheeran, B. and Tauscher, G. and Wesch, S. and Wing, M. and Winkler, P. and Zeng, M. and Osterhoff, J.},
}

@conference{ben,
    author = {J. B. B. Chen and others},
    title = {{ABEL: The adaptable beginning-to-end linac simulation framework}},
    pages = {},
    paper = {TUPS012},
    intype = {presented at the},
    series = {International Particle Accelerator Conference},
    publisher = {JACoW Publishing, Geneva, Switzerland},
    note = {presented at IPAC'25, Taipei, Taiwan, Jun. 2025, paper TUPS012, this conference},
    language = {english}
}

@article{Lindstrom,
issn = {1748-0221},
journal = {Journal of instrumentation},
pages = {P05016},
volume = {17},
publisher = {IOP Publishing},
number = {5},
year = {2022},
title = {Emittance preservation in advanced accelerators},
copyright = {2022 The Author(s)},
language = {eng},
address = {Bristol},
author = {Lindstrøm, C.A. and Thévenet, M.},
}

@article{Lee,
    author = {Lee, S. and Katsouleas, T.},
    title = {Wakefield accelerators in the blowout regime with mobile ions},
    journal = {AIP Conference Proceedings},
    volume = {472},
    number = {1},
    pages = {524-533},
    year = {1999},
    month = {07},
    issn = {0094-243X},
    doi = {10.1063/1.58913},
    url = {https://doi.org/10.1063/1.58913},
    eprint = {https://pubs.aip.org/aip/acp/article-pdf/472/1/524/12044261/524\_1\_online.pdf},
}

@article{Gholizadeh,
    author = {Gholizadeh, Reza and Katsouleas, Tom and Muggli, Patric and Mori, Warren},
    title = {Analysis of Ion Motion and Scattering in the Extreme Regime of High Intensity Electron Beams in Plasma Wakefield Accelerators},
    journal = {AIP Conference Proceedings},
    volume = {877},
    number = {1},
    pages = {504-510},
    year = {2006},
    month = {11},
    issn = {0094-243X},
    doi = {10.1063/1.2409176},
    url = {https://doi.org/10.1063/1.2409176},
    eprint = {https://pubs.aip.org/aip/acp/article-pdf/877/1/504/11908848/504\_1\_online.pdf},
}

@article{uniform-ramp,
  title = {Transverse beam dynamics in a plasma density ramp},
  author = {Ariniello, R. and Doss, C. E. and Hunt-Stone, K. and Cary, J. R. and Litos, M. D.},
  journal = {Phys. Rev. Accel. Beams},
  volume = {22},
  issue = {4},
  pages = {041304},
  numpages = {11},
  year = {2019},
  month = {Apr},
  publisher = {American Physical Society},
  doi = {10.1103/PhysRevAccelBeams.22.041304},
  url = {https://link.aps.org/doi/10.1103/PhysRevAccelBeams.22.041304}
}
	{
}
\end{document}